\documentclass[final,3p,times,twocolumn]{elsarticle}
\usepackage{graphics}
\usepackage{epsfig}
\usepackage{amssymb}

\journal{International Journal of Mass Spectrometry}

\begin{document}

\begin{frontmatter}

\title{A Broad-Band FT-ICR Penning Trap System for KATRIN \tnoteref{label1}}
\tnotetext[label1]{The results which are presented here are part of the PhD
  thesis of M.~Ubieto-D\'iaz}

\author[label2]{M.~Ubieto-D\'iaz}
\author[label3]{D. Rodr\'iguez\corref{cor1}}
\cortext[cor1]{Corresponding author} 
\ead{danielrodriguez@ugr.es}
\author[label4]{S.~Lukic}
\author[label2,label5]{Sz.~Nagy}
\author[label6]{S.~Stahl}
\author[label2]{K.~Blaum}

\address[label2]{Max-Planck-Institut f\"ur Kernphysik, 69117 Heidelberg, Germany}
\address[label3]{Universidad de Granada, 18071, Granada, Spain}
\address[label4]{University of Karlsruhe, Institute for Experimental Nuclear Physics, 76344 Eggenstein-Leopoldshafen, Germany}
\address[label5]{EMMI GSI Helmholtzzentrum f\"ur Schwerionenforschung, 64291
  Darmstadt, Germany}
\address[label6]{Stahl-Electronics, Kellerweg 23, 67582 Mettenheim, Germany}

\begin{abstract}
The KArlsruhe TRItium Neutrino experiment KATRIN aims at improving the upper
limit of the mass of the electron antineutrino to about 0.2 eV (90$\%$ c.l.) 
by investigating the $\beta$-decay of tritium gas molecules
$\hbox{T}_{2}\rightarrow
(^{3}\hbox{HeT})^{\scriptsize{+}}+e^{\scriptsize{-}}+\bar \nu
_{\hbox{\scriptsize{e}}} $. The experiment is currently under construction to
start first data taking in 2012. One source of systematic uncertainties in the
KATRIN experiment is the formation of ion clusters when tritium decays and
decay products interact with residual tritium molecules. It is essential to
monitor the abundances of these clusters since they have
different final state energies than tritium ions. 
For this purpose, a prototype of a cylindrical Penning trap
has been constructed and tested at the Max-Planck-Institute for Nuclear
Physics in Heidelberg, which will be installed in the KATRIN beam line. This
system employs the technique of Fourier-Transform Ion-Cyclotron-Resonance
in order to measure the abundances of the different stored ion 
species.

\end{abstract}


\end{frontmatter}


\section{Introduction}

The observation of neutrino oscillations in recent experiments implies a
non-zero neutrino mass ($m_{\scriptsize{\nu }}\neq 0$) \cite{AHM02}. So far, the
experiments aiming at directly measuring the neutrino mass only succeeded to
set an upper limit (see e.g. the recent review by Otten and Weinheimer \cite{OTE08}). 

\begin{figure*}[t]

  \includegraphics[scale=0.6,angle=270]{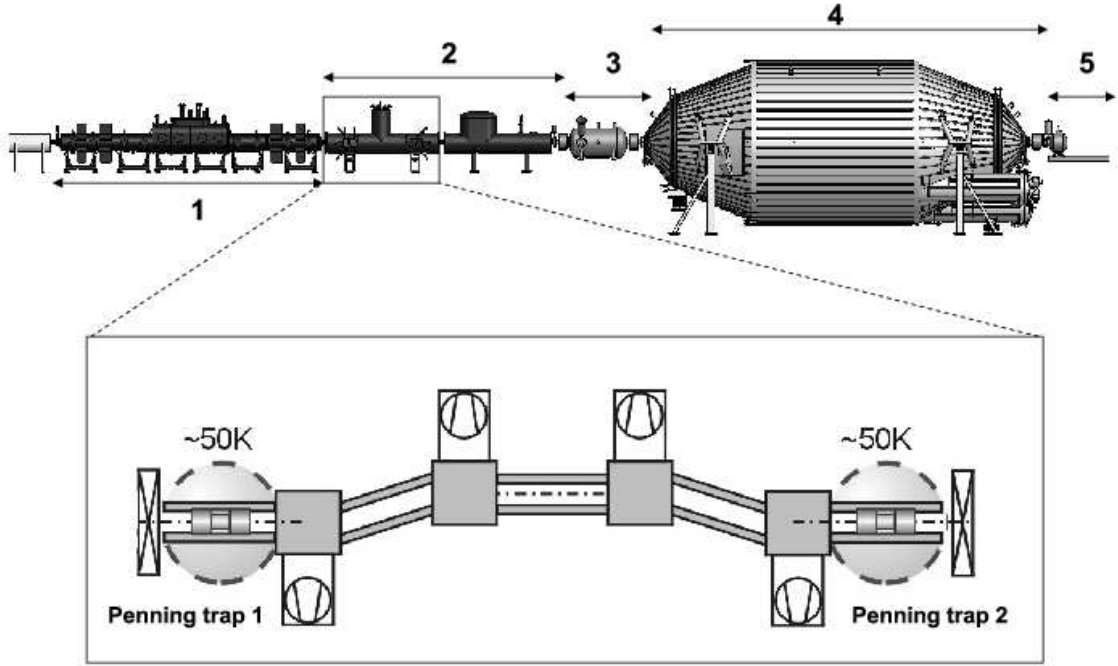}
\vspace{-2cm}  
\caption{Sketch of the KATRIN beam line indicating the
  main components: 1) windowless gaseous tritium source, 2) transport
  section, 3) pre-spectrometer, 4) main spectrometer, and 5) detector. The
  total length of the KATRIN beam line is 70~m. In the first part of the
  transport section (enlarged at the bottom) the concentration of
  unwanted ions due to the formation of clusters will be monitored employing 
  Penning traps using the Fourier-Transform Ion-Cyclotron-Resonance (FT-ICR)
  detection technique. The traps will be operated at cryogenic temperatures in
  a magnetic field of 5.6~T.} 
\label{fig:1}   
\end{figure*}

The most recent evaluation of the limit of the antineutrino mass, which yields $m_{\bar \nu }\leq 2.3$~eV (90$\%$ c.l.) \cite{KRA05}, is based on the results
from two experiments, both performed by means of an electrostatic
spectrometer, one in Troitsk \cite{LEBE95} and the other at the University of
Mainz \cite{KRA05}. In these experiments, the neutrino  mass is extracted by
analyzing the energy region near the end-point energy $E_{0}$ of the
$\beta$-decay spectrum of tritium ($T_{1/2}=12.3$~y, $E_{0}=18.589 8(12)$~keV
which is derived from the $^{3}$He-T mass difference
\cite{NAG06}). In the decay $\hbox{T}_{2}\rightarrow
(^{3}\hbox{HeT})^{\scriptsize{+}}+e^{\scriptsize{-}}+\bar \nu
_{\hbox{\scriptsize{e}}}$, only an energy interval $\delta
E=10$~eV below $E_{0}$ is used for the analysis. However, the fraction of
electrons in this energy interval is only $2\times 10^{-10}$. In order to
improve the sensitivity or reduce the upper limit of  $m _{\bar
  \nu }$ by studying the $\beta$-decay of T$_2$, long data collection periods 
using a strong tritium source and a spectrometer with a large acceptance solid 
angle for the electrons coming from the decay are required. To this end, the KArlsruhe 
TRItium Neutrino experiment (KATRIN) \cite{ANG04} was initiated
aiming at reaching $m _{\bar \nu }\leq 0.2$~eV (90$\%$
c.l.). Figure~\ref{fig:1} shows the KATRIN beam line indicating the main components.

An important issue to control systematic effects in KATRIN is the precise
knowledge of the abundance of ion species such as T$^{+}$, T$^{-}$,
$^{3}$He$^{+}$, ($^{3}$HeT)$^{+}$, T$_2$$^{+}$, T$_3$$^{+}$, T$_5$$^{+}$, and
even larger cluster ions formed by $\beta$ decay, ionization processes and
chemical reactions, since they have different final state energy spectra than
$\hbox{T}_{2}$. Assuming a T$_2$ density of $5 \cdot 10^{14}$~cm$^{-3}$ in the
source position, the expected density for T$^{-}$ is
$\leq 10^{7}$~cm$^{-3}$, and the one for T$^{+}$ is in the range
$10^{4}$-$10^{5}$~cm$^{-3}$. For monitoring these ion species, two Penning trap systems will
be installed in the transport section of KATRIN. This section is marked with $\#$2
in the upper part of Fig.~\ref{fig:1}. The comparison of the
concentrations of the different species observed in the
two traps will also serve to monitor the level of suppression of contaminating
ions by the use of crossed electric and magnetic fields in the cryogenic
transport section, while allowing the $\beta $-decay electrons to pass and to enter the spectrometers. 

The monitoring of the ion species captured in the Penning trap is
done by detecting the image current induced in the ring electrodes. The
mass-to-charge ratio of the different ion species confined in 
the trap is extracted from the measurements of their so-called
modified-cyclotron frequencies  ($\nu _{\scriptsize{+}}$). This 
is a non-destructive detection method known as Fourier-Transform
Ion-Cyclotron-Resonance (FT-ICR) technique \cite{COM78,MAR98},
commonly used in chemistry \cite{MAR00}. The ratio of the different cluster 
ions formed in the KATRIN source can be determined since the FT-ICR
technique allows monitoring of all the ion species simultaneously. The 
amplitudes of the signals observed in the frequency spectrum are
proportional to the number of ions for each ion species. 

The maximum resolving power $\nu _{\scriptsize{+}}/\Delta \nu
_{\scriptsize{+}}$ required for the FT-ICR Penning traps to be 
installed at KATRIN is in the order of $1\times 10^{5}$ for
$\nu _{\scriptsize{+}}\sim 20$~MHz in order to distinguish 
T$^{+}$ from $^{3}$He$^{+}$ ions. The minimum number of ions needed to observe an
FT-ICR signal is expected to be in the order of a few thousand ions in 
the mass range of interest. This sensitivity will be further enhanced once the
traps and the detection system are installed in the KATRIN beam line at the
cryogenic temperature $T=50$~K. In the following, the KATRIN Penning trap
system as well as its commissioning will be presented. For the characterization, the mass-to-charge ratios of the different test species were not monitored simultaneously in the frequency spectrum. 

\section{The KATRIN Penning trap: principle and experimental setup}

In the ideal Penning trap, the electric potential is harmonic and
the ions are confined by the combination of an electrostatic and a magnetic
field \cite{GAB86,BLA06}. 
\begin{figure*}[t]
\resizebox{0.85\textwidth}{!}{%
  \includegraphics[angle=270]{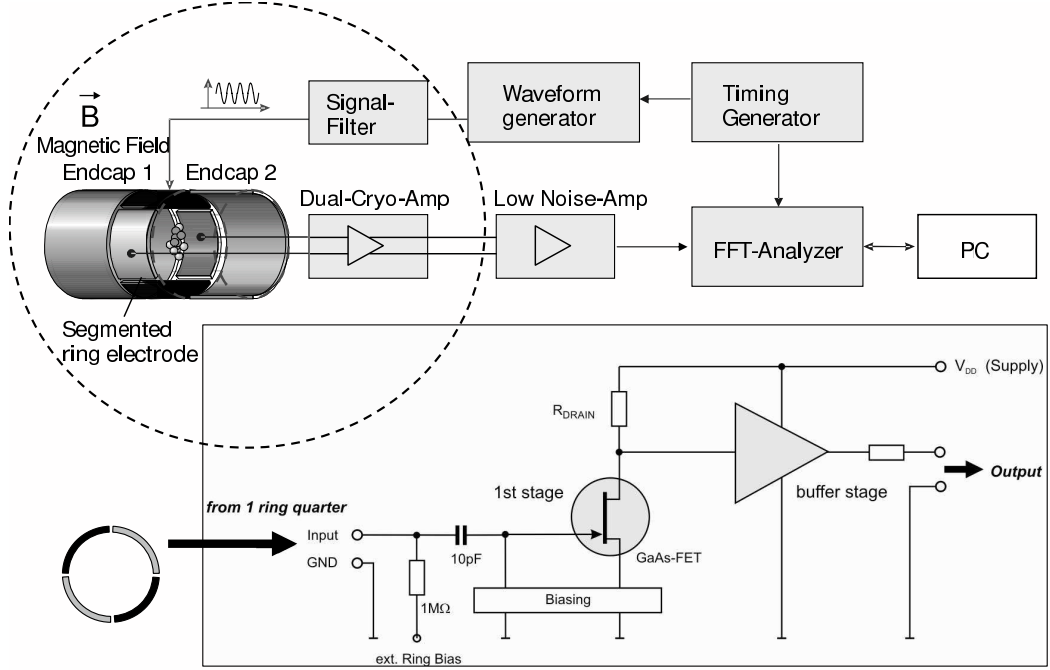}
}
\vspace{-40pt}  
\caption{Schematic view of the three-electrode cyclindrical Penning trap system to be
  placed in the transport section of the KATRIN beam line. The dashed circle
  houses all the components: Penning trap, cryogenic preamplifier and signal
  filter, located in vacuum in the high-magnetic field region of a
  superconducting magnet. The waveform generator used for excitation provides
  also the trigger for data acquisition with the FFT analyzer. The square box
  shows a simplified version of the electronic circuit for one channel of the
  Dual-Cryo-Amplifier.}
\label{fig:3}   
\end{figure*} 
This electric field allows for confinement in the axial direction ($\vec B
\parallel \vec z$), while the magnetic field confines the ions in the radial plane. As
a result, the motion of trapped charged particles in an ideal Penning trap can
be described as the superposition of three independent motions with
characteristic eigenfrequencies. The frequencies are:
\begin{equation}
\nu _{\hbox{\scriptsize{z}}}=\frac{1}{2\pi}\sqrt{\frac {qU}{md^{2}}},\label{axial}
\end{equation}
for the axial motion and 
\begin{equation}
\nu _{\scriptsize{\pm}}=\frac{\nu _{\hbox{\scriptsize{c}}}}{2}\pm\sqrt {\frac{\nu
    _{\hbox{\scriptsize{c}}}^{2}}{4}-\frac {\nu _{\hbox{\scriptsize{z}}}^{2}}{2}},\label{reduced}
\end{equation}
for the modified cylotron (+) and magnetron ($-$) motions, respectively. 
In Eq.~(\ref{axial}), $m$ is the mass of the ion, $q$ the electric
charge, $U$ the potential difference between ring and endcap electrodes, and
$d$ is a parameter which describes the trap dimensions \cite{GAB86}. This
parameter reads:
\begin{equation}
d=\sqrt{\frac{z_{\scriptsize{0}}^{\scriptsize{2}}}{2}+\frac{r_{\scriptsize{0}}^{\scriptsize{2}}}{4}},
\end{equation}
where $r_{\scriptsize{0}}$ and $z_{\scriptsize{0}}$ are the minimum distance
from the center of a hyperbolic trap to the ring and endcap electrodes,
respectively. $\nu _{\hbox{\scriptsize{c}}}$ in Eq.~(\ref{reduced}) represents the cyclotron
frequency of a charged particle moving in a pure magnetic field and is given by:
\begin{equation}
\nu _{\hbox{\scriptsize{c}}}=\frac {1}{2\pi }\cdot \frac{qB}{m}. \label{cyclotron} \end{equation}
The characteristic frequency of the reduced-cyclotron motion is slightly lower
than the free cyclotron frequency. In an ideal Penning trap
Eq.~(\ref{reduced}) obeys the relation $\nu _{\hbox{\scriptsize{c}}}=\nu
_{\hbox{\scriptsize{+}}}+\nu _{\scriptsize{-}}.\label{relation}$ In a real
Penning trap, $\nu _{\hbox{\scriptsize{c}}}^{2}= \nu
_{\hbox{\scriptsize{+}}}^{2}+\nu _{\hbox{\scriptsize{-}}}^{2}+\nu
_{\hbox{\scriptsize{z}}}^{2}$ is valid and it is known as the Brown-Gabrielse
invariance theorem \cite{BRO82}.

The identification of the ions in FT-ICR mass spectrometry is mainly done by
measuring $\nu _{\hbox{\scriptsize{+}}}$ \cite{MAR98}. The induced ion current
is measured during a fixed time and the frequency is unfolded after applying
Fourier transformation. The induced current of an ion with kinetic energy $E$
is given by the relationship \cite{COM78}:
\begin{equation}
I_{\hbox{\scriptsize{eff}}}=\frac {q}{D}\cdot \sqrt {\frac {E}{m}}, \label{eq:1}
\end{equation}
where $D$ is the so-called effective distance in the trap, which is in the
order of the trap diameter. This current, which is typically in the order of a
few fA, transforms into a voltage signal by flowing through parasitic
capacitances, either of the trap itself or of the attached current
amplifiers. The resulting voltage signal for one ion is
$I_{\hbox{\scriptsize{eff}}}/(\omega C)$, where $C$ is the sum of all
capacitances present at the trap electrodes. In the KATRIN trap $C \sim 25$~pF.

The KATRIN trap system is a three-electrode cylindrical Penning trap \cite{GAB86} shown
schematically in Fig.~\ref{fig:3}. The trap  with an inner diameter of $\Phi =
71$~mm consists of one ring and two endcap electrodes. This is
unusually large in order to allow for an efficient transport of the electrons from the
source to the spectrometers in the KATRIN experiment. 
The ring electrode is four-fold segmented as shown schematically in
Fig.~\ref{fig:3}. One pair of opposite segments is used for excitation of the
ion motion and the other one is used for detection. The excitation is done
by applying a radio-frequency (RF) dipolar field scanning the frequency around
the modified cyclotron frequency of the ions stored in the trap. The ions then
gain energy from the RF dipolar field and the induced current
$I_{\hbox{\scriptsize{eff}}}$ is detectable provided a) sufficient number of ions
are stored in the trap, b) they move coherently, and c) the resulting signal
is above the noise level. The signal-to-noise ratio is given by:
\begin{equation}
S/N=\frac{1}{(BCD)\cdot n_{\hbox{\scriptsize{d}}}}\cdot \sqrt{\frac{E\cdot m}{B_{\hbox{\scriptsize{FFT}}}}}
\end{equation}
where $B,C,D$ are introduced above, $B_{\hbox{\scriptsize{FFT}}}$ the
bandwidth of one FFT channel, and $n_{\hbox{\scriptsize{d}}}$ is the noise
density. The later is a measurement of the input voltage noise in a certain
frequency width at the analyzer and is given in V/Hz$^{1/2}$.

\begin{figure*}[t]
\begin{center}
\resizebox{0.9\textwidth}{!}{%
  \includegraphics[angle=270]{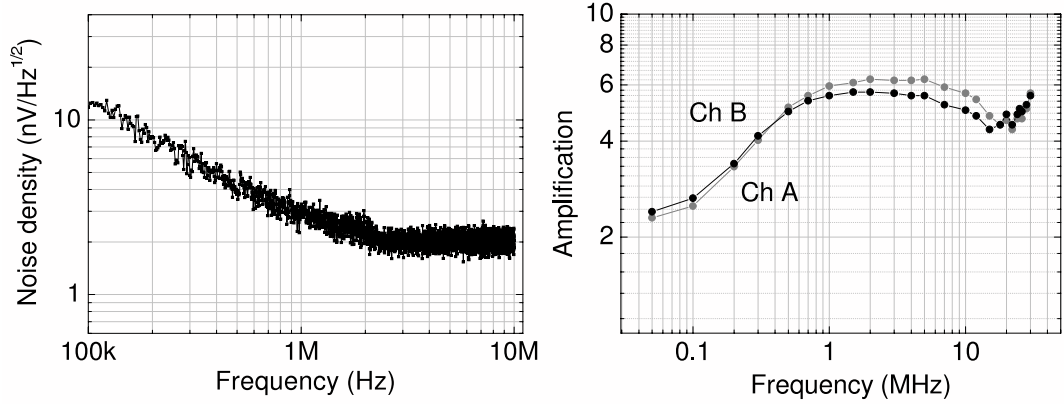}
}  
\vspace{-40pt}
\caption{Noise level and amplification factor versus frequency of the
  amplifier designed to be operated at cryogenic
  temperatures in the KATRIN beamline. The noise level was measured at
  $T=300$~K and is expected to be reduced by a factor of
  $\approx 1.5$ at $T=50$~K. For the determination of the amplification factor, 
  the signals from the two quarters of the ring were fed into the measuring
  device through different channels (ChA and ChB).}
\label{fig:4}   
\end{center}  
\end{figure*}

Amplifier boxes containing the preamplification stages, which are designed to
be operated at a later time under cryogenic temperatures, are attached to the
trap. In a first stage the signal is passed through a GaAs Field Effect
Transistor (FET) amplifier mounted on the trap assembly. The electronic
circuit for one channel is shown in the square box of Fig.~\ref{fig:3}. Some
characteristics of the amplifier are shown in Fig.~\ref{fig:4}. In the
frequency range of interest (1-20~MHz) a noise-density level below
3~nV/Hz$^{1/2}$ and an amplification of about $6$ was achieved. The second
amplification (25-fold amplification) takes place outside the vacuum
chamber. The signal is fed into an FFT analyzer to compute the frequency
spectrum. For the measurements presented here an analyzer from Rohde $\&$
Schwarz model FSP3 with a frequency range from 9~kHz to 3~GHz was used.

The test setup comprises an electron impact ion source, a micro-channel plate
detector and the Penning trap system housed in a
superconducting magnet. The magnet provides a field strength of 4.7~T in
the center of the trap. This value is slightly smaller than the field strength
in the KATRIN beam line (5.6~T). For the tests presented here ions were
created either by discharges or by electron impact on gas molecules and atoms
inside the trap. The electrons were produced using an ion source with
a photoelectric-emission cathode constructed to test different 
components of the KATRIN experiment \cite{SCHO08}. 

\section{Results}
\begin{figure}[t]
\begin{center}
\resizebox{0.5\textwidth}{!}{%
  \includegraphics[angle=270]{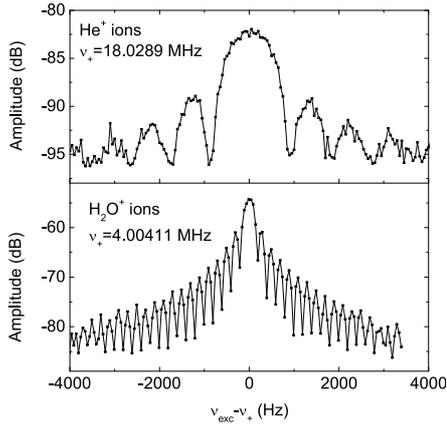}
}  
\caption{Top: amplitude of the FT-ICR signal versus excitation frequency ($\nu_{\hbox{\scriptsize{exc}}}$) for He$^{+}$ ions ($\Delta \nu
  _{\hbox{\scriptsize{FWHM}}}=1095$~Hz). Bottom: amplitude of the FT-ICR
  signal versus excitation frequency for
  H$_{2}$O$^{+}$ ions ($\Delta \nu
  _{\hbox{\scriptsize{FWHM}}}=276$~Hz). $\nu_{\hbox{\scriptsize{+}}}$ is the
  modified cyclotron frequency. For details see text.}
\label{fig:5}
\end{center}
\end{figure}

The measurements presented here were performed at room temperature
using  He$^{+}$, H$_2$O$^{+}$ and N$_2$$^{+}$ ions as test species. All species
were identified by exciting the ion motion at their modified cyclotron
frequency. The trap system was running in repeated cycles. The cycle time was 100~ms and the measurement sequence consisted of three steps:
\begin{itemize}
\item{injection (100~$\mu$s): endcap $1$ at 0~V and endcap $2$ at potential $U$.}
\item{trapping (100~ms): endcap $1$ and $2$ at potential $U$.}
\item{ejection (100~$\mu$s): endcap $1$ at potential $U$ and endcap $2$ at 0~V.}
\end{itemize}
The ring electrode was always at ground potential. During trapping, dipolar
excitation at a frequency $\nu _{\scriptsize{\hbox{rf}}}$ around the nominal $\nu
_{\scriptsize{+}}$ value was applied. This excitation frequency was varied after each
cycle ``injection-trapping-ejection'' in steps of 50~Hz. The amplitude of the
induced (FT-ICR) signal was recorded every cycle for the corresponding $\nu
_{\scriptsize{\hbox{rf}}}$ frequency. Figure~\ref{fig:5} shows the amplitudes
of the FT-ICR signals for trapped He$^{+}$ and H$_{2}$O$^{+}$ ions. The number
of RF excitation periods in the waveform generator (AGILENT model 33250A) was
20,000. Thus, the duration for the excitation for any cycle (data point) was $\sim 1.1$~ms for He$^{+}$
ions and $\sim 5$~ms for H$_{2}$O$^{+}$ ions.
The side bands in Fig.~\ref{fig:5} arise from the finite excitation, which has
in the time domain a rectangular shape. The mass resolving power $m/\Delta
m$ with $\Delta m$ being the full-width-at-half-maximum is $1.6 \times 10^{4}$
for He$^{+}$ ions and $1.5 \times 10^{4}$ for H$_{2}$O$^{+}$
ions, respectively. Measurements with He$^{+}$ ions were also carried out for
longer excitation times of 5.5~ms using 100,000~RF~periods resulting in
$m/\Delta m=7.2\times 10^{4}$. A resolving power of $1.6\times 10^{5}$ will be
required to distinguish in the frequency spectrum T$^{+}$ from $^{3}$He$^{+}$
ions. This will be possible due to the higher frequency and by chosing a
proper excitation time of about 10~ms. For all these measurements the 
acquisition time to record the image current was $5$~ms. Larger
observation times imply a reduction in the amplitude of the FT-ICR signal 
due to the short coherence time. 

\begin{figure}[t]
\begin{center}
\resizebox{0.5\textwidth}{!}{%
  \includegraphics{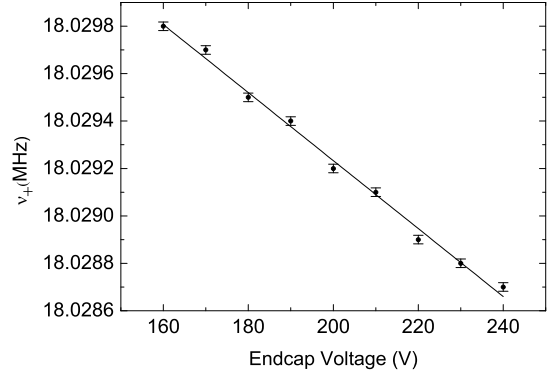}
}  
\caption{Modified cyclotron frequency ($\nu _{\hbox{\scriptsize{+}}}$) for
  He$^{+}$ ions as a function of the
  voltage applied to the endcap electrodes. A relative uncertainty of
  $10^{-6}$ due to magnetic field inhomogeneities was added to the
  modified cyclotron frequencies.}
\label{fig:6}   
\end{center}  
\end{figure}

After probing the excitation, the FT-ICR signal at $\nu
_{\scriptsize{\hbox{rf}}}=\nu _{\scriptsize{\hbox{+}}}$ was recorded for different
potentials $U$ applied to the endcap electrodes. Figure~\ref{fig:6} shows the
modified cyclotron frequency for He$^{+}$ ions as a function of the voltage applied to the
endcap electrodes. The solid line is the result of a least-square linear 
fit following the relationship:
\begin{equation}
\nu _{\scriptsize{+}}=\frac {qB}{2\pi m}-\frac{U}{4\pi d^{2}B},\label{fit}
\end{equation}
which is obtained from Eq.~(\ref{relation}) by substituting $\nu
_{\hbox{\scriptsize{c}}}$ and $\nu _{\scriptsize{-}}$. Applying a linear fit to the
data shown in Fig.~\ref{fig:6} and using the mass value for He$^{+}$
\cite{AME03}, one can obtain the magnetic field from the
first term in Eq.~(\ref{fit}) as $B=4.6995(2)$~T. Similar measurements were
perfomed with H$_{2}$O$^{+}$ ions yielding $B=4.6996(6)$~T. 
The parameter $d$ was also obtained using Eq.~(\ref{fit}). The results are $d$(He)$=3.43(2)$~cm and
$d$(H$_{2}$O)$=3.48(2)$~cm, also in excellent agreement.

\begin{figure}[t]
\begin{center}
\resizebox{0.5\textwidth}{!}{%
  \includegraphics{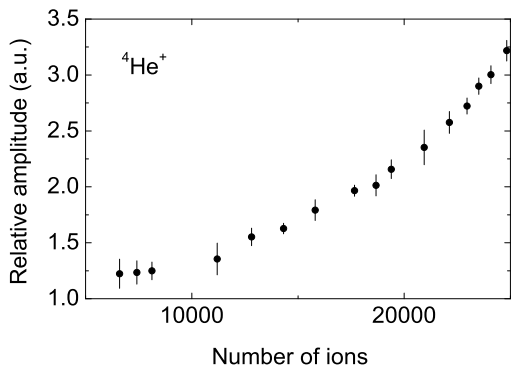}
}  
\caption{Signal-to-noise ratio versus number of ions extracted from the
  trap. Saturation of the micro-channel plate detector is observed for ion
  numbers above 20,000 since the ions are ejected from the trap in a pulsed
  mode. The system was tuned so that only He$^{+}$ ions were observed in the
  time-of-flight spectrum recorded with the micro-channel plate.}
\label{fig:7}
\end{center}   
\end{figure}

The last set of measurements presented here to characterize the Penning traps
was dedicated to determine the minimum number of ions needed to observe an
FT-ICR signal at room temperature. For these measurements a set of two attenuation grids were
mounted in front of the micro-channel plate detector resulting in an intensity reduction by a
factor of $100$. This allowed the characterization of the trap with up to about 20,000
ions. The attenuation factor of 100 was measured with laser light and a
photodiode. The voltage signal from the photodiode was measured using a 
multimeter for both situations, i.e. with and without the attenuation grids behind the laser beam. 
The results of the measurements are shown in Fig.~\ref{fig:7} for He$^{+}$
ions. The $y$-axis of the figure is the signal-to-noise ratio recorded with the FFT
analyzer running in average mode. Each data point is the average over 100 measurements. 
The number of detected ions was corrected for the micro-channel plate detection
efficiency of 30$(5)\%$ \cite{WIZ79}. After correction, the minimum number for
$^{4}$He$^{+}$ ions is around 6,000. This number is higher than the one for
H$_{2}$O$^{+}$ ions, which was measured to be around 1,000. This can be
explained by the different effective voltages ($U_{\hbox{\scriptsize{eff}}}
\propto \sqrt {m}$) and by the different amplification factors in the
frequency domain (see Fig.~\ref{fig:4}). 


The maximum number of ions detectable by this technique was not yet reached 
in the experiments described above. The maximum FT-ICR signal measured yielded
an amplitude four times higher than the one associated to 20,000 trapped ions
shown in Fig.~\ref{fig:7}. The number of ions arriving at the micro-channel plate was not
measured due to saturation of the detector. The characterization in this case
requires further attenuation in front of the detector. If one assumes a linear
dependence, the maximum signal observed would correspond to about 150,000
He$^{+}$ ions. FT-ICR signals with larger amplitudes have been observed for H$_2$O$^+$ ions.    

\section{Conclusions and outlook}
In this paper, the experimental setup as well as first results obtained with
the three-electrode Penning trap system developed for the KATRIN experiment are
presented. Excitation at the modified cyclotron frequency of stored ions and
detection of the induced image current in the trap electrodes have been
accomplished. These results show the proof of principle of the three-electrode
Penning trap system to be operated in the KATRIN beamline and reveal its present
performance. The noise-signal density at room temperature of
the GaAs FET-based amplifier is $<3$~nV/Hz$^{1/2}$, and the
amplification factor is between 4 and 6 in the frequency range of interest for
KATRIN. In this range, the present limit in sensitivity is around 6,000 ions
for mass $A=4$ and about 1,000 ions for $A=18$. Still the sensitivity limit can be improved once
the three-electrode Penning trap system is operated at cryogenic temperatures in
the transport section of the KATRIN beam line. The resolving power has to
exceed $1.6\times 10^{5}$ to distinguish in the frequency spectrum T$^{+}$ from $^{3}$He$^{+}$ ions. This is feasible with the current system if excitation times above 10~ms are used. 

\section{Acknowledgement}
The two Penning traps have been financed by the BMBF (grant to the University
of Karlsruhe) under project codes 05CK5VKA/5 and 05A08VK2. The support of the
Deutsche Forschungsgemeinschaft for the development of the FT-ICR detection
technique for precision mass spectrometry under contract 
number BL981-2-1 is gratefully acknowledged. We thank A. Gotsova for her help
during tests in Mainz and Prof. C. Weinheimer for useful discussions related
to this project. We warmly thank the LPC trappers group for providing the
attenuation grids. D.~Rodr\'iguez is a Juan de la Cierva fellow and
acknowledges support from the Spanish Ministry of Science and Innovation
through the Jos\'e Castillejo program to provide funding for a five-months
stay at the MPI-K. Sz.~Nagy acknowledges support from the Alliance Program of
the Helmholtz Association EMMI. S.~Luckic acknowledges support by the Transregional Collaborative Research Centre No. 27 ``Neutrinos and Beyond'', funded by Deutsche Forschungsgemeinschaft.

\end{document}